\pdfoutput=1

\documentclass{article}
\usepackage{spconf,amsmath,graphicx}

\newcommand\norm[1]{\left\lVert#1\right\rVert}

\usepackage{amsfonts}
\usepackage{booktabs}
\usepackage{multirow}

\title{LOW-LATENCY DEEP CLUSTERING FOR SPEECH SEPARATION}
\name{Shanshan Wang\thanks{The authors wish to thank CSC-IT Centre of Science Ltd., Finland,  for providing computational resources used in  experiments reported in this paper.}, Gaurav Naithani, Tuomas Virtanen \vspace{0mm}}
\address{Laboratory of Signal Processing, Tampere University of Technology, Tampere, Finland \\ \vspace{0mm}
 \normalsize Email:\{ shanshan.2.wang, gaurav.naithani, tuomas.virtanen\}@tut.fi
}
\begin{document}
%
\maketitle
\vspace{0mm}
\begin{abstract}
 This paper proposes a low algorithmic latency adaptation of the deep clustering approach to speaker-independent speech separation. It consists of three  parts: \mbox{ \textit{a)}} the usage of long-short-term-memory (LSTM) networks instead of their bidirectional variant  used in the original work, \mbox{\textit{b)} } using a short synthesis window (here 8 ms) required for low-latency operation, and, \mbox{\textit{c)} using} a buffer in the beginning of  audio mixture to estimate cluster centres corresponding to constituent speakers which are then utilized to separate speakers within the rest of the signal. The buffer duration would serve as an initialization phase after which the system is capable of operating with 8 ms algorithmic latency. We evaluate our proposed approach on two-speaker mixtures from  Wall Street Journal (WSJ0) corpus. We observe that the use of LSTM yields around one dB lower SDR as compared to the  baseline bidirectional LSTM in terms of source to distortion \mbox{ratio~(SDR)}. Moreover, using an 8 ms synthesis window instead of 32 ms degrades the separation performance by around 2.1 dB as compared to the baseline. Finally,  we also report separation performance with different buffer durations noting that separation can be achieved even for buffer duration as low as 300 ms. 



\end{abstract}
\vspace{0mm}
\section{Introduction}
\label{sec:intro}
Single channel speech separation is the problem of recovering the constituent speech signals from an acoustic mixture signal when information from only a single microphone is available \cite{vincent2018audio}. In recent years, data-driven methods relying on deep neural networks (DNN) \cite{huang2014deep, erdogan2015phase} have yielded dramatic improvements in performance in comparison to the previously used methods, e.g., model-based approaches \cite{roweis2001one}) and  matrix factorization \cite{virtanen2007monaural, schmidt2006single}. In particular, speaker-independent speech separation has been addressed by approaches like  deep clustering  \cite{hershey2016deep, Isik+2016}, permutation invariant training \cite{yu2017permutation}, and more recently deep attractor networks \cite{chen2017deep}  which is the current state-of-the-art. Improvements to the original deep clustering framework \cite{hershey2016deep} have been proposed in terms of, e.g., better objective functions \cite{wang2018alternative}; and, improved regularization and curriculum training \cite{Isik+2016}. These studies have considered offline separation scenario where the signal to be separated is available at once.

Low-latency processing is important when these DNN-based methods are applied to applications like hearing aids \cite{bramslow2010preferred} and cochlear implants \cite{hidalgo2012low}, In particular, for hearing aids,  the latency requirements are quite restrictive as the sound is perceived by the listener via hearing aid as well as the direct path. Several studies have documented the subjective disturbance  experienced by the listeners (e.g., \cite{stone2008tolerable}). Notably,  Agnew~{\textit{et~al}}.~\cite{agnew2000hearing}  found the delays above \mbox{10 ms} to be objectionable while  delays as low as 3 to \mbox{5 ms} to be noticeable by hearing-impaired listeners. 

For the above applications, the offline DNN-based methods run into two main problems. Firstly, we do not have access to the future temporal information hence DNN models like bidirectional long-short-term-memory networks (BLSTM), as used in  \cite{erdogan2015phase, chen2015speech, hershey2016deep, Isik+2016}, cannot be used. Secondly, for short-time Fourier transform (STFT) based  systems, the algorithmic latency is at least equal to  the frame length of synthesis window used for signal reconstruction. This limits us from using window sizes used  in conventional speech processing ( e.g., 20 -40 ms \cite{paliwal}).  Speech separation methods with algorithmic latencies below \mbox{10 ms} have been reported, e.g., using non-negative matrix factorization \cite{barker2015low}, and DNNs \cite{naithani2017low, naithani2016low, tasnet2018}.


In this paper, we investigate a low-latency adaption of the deep clustering framework first introduced in \cite{hershey2016deep}. The original framework involves using a BLSTM network to estimate high-dimensional embeddings for each time-frequency bin in the mixture STFT which is then partitioned into clusters corresponding to the constituent speakers. Our focus in this work is three-fold: \textit{a)} investigation of separation performance with LSTM instead of BLSTM to allow online processing,  \textit{b)} investigation of separation performance for short synthesis window (8 ms in this work) instead of longer 32 ms used in original work, and \textit{c)} investigation of using a certain duration in the beginning of acoustic mixture to estimate the cluster centres corresponding to the constituent sources. We refer to this time duration as the buffer. The estimation of cluster centres here thus serves as an initialization phase and the method is capable of doing online separation  after the buffer duration. 


We evaluate separately the effect of the above three modifications. We observe  one dB lower SDR while using an LSTM instead of the  BLSTM network  as was used in the original work \cite{hershey2016deep}. The separation performance  degrades  by around 2.1 dB while shortening the window length from 32 ms to 8 ms as compared to the baseline. Moreover, we show that it is possible to estimate reasonably well cluster centres using just the beginning of the signal yielding good separation for the rest of the signal. 


The paper is structured as follows: Section \ref{sec:method} describes the baseline deep clustering approach proposed in \cite{hershey2016deep}. Section \ref{sec:low_latency} describes its low-latency adaptation. Section \ref{sec:evaluation} describes the evaluation procedure, experimental set up, and obtained results. Finally, Section \ref{sec:conclusion} concludes the paper.

\vspace{1ex}

\section{Deep Clustering for Speaker Separation}\label{sec:method}

In this section, we summarize the deep clustering method proposed in \cite{hershey2016deep, Isik+2016}. Deep clustering  can be thought of as a combination of supervised learning and unsupervised learning. Unlike the traditional DNN-based speech separation methods that predict a time-frequency mask or separate speech spectrum for the mixture input in a supervised manner \cite{huang2014deep, erdogan2015phase}, it generates an embedding vector for each time-frequency bin and then uses the unsupervised learning approach such as k-means to cluster the embedding vectors in order to get the time-frequency masks.

Given a mixture audio signal in the time domain $x(n)$, firstly, features are extracted  by calculating its log magnitude short-time Fourier transform (STFT).  The features are then inputted to a neural network that will output an embedding vector for each of the time-frequency points. In the original deep clustering framework, BLSTM network was used \cite{hershey2016deep}, and therefore we choose it as the baseline here. The output of the neural networks is an embedding matrix $\mathbf{V}\mathbf{\in}\mathbb{R}^{T F \times D}$, where $T$ denotes the number of frames, $F$ the number of frequency bins, and $D$ the embedding dimension. Finally, k-means clustering is employed to partition the embedding vectors into clusters corresponding to different constituent speakers. Binary time-frequency masks for each speaker is then obtained using these cluster assignments by assigning 1 to all the time-frequency bins within the cluster of the speaker, and 0 to the rest of the bins.




The neural network is trained to minimize the difference between the estimated affinity matrix $\mathbf{VV}^\mathcal{T}$  derived from the embeddings $\mathbf{V}$ predicted by the neural network and the target affinity matrix $\mathbf{YY}^\mathcal{T}$, where  $\mathbf{Y}\in $ $\mathbb{R}^{T F \times C}$  is the ideal binary mask. $C$ indicates the number of speakers in the mixture. The training loss function $\mathcal{L} $ is computed as,
\vspace{-2ex}

\begin{align} \label{eq:loss}
\mathcal{L} &=  \norm{\mathbf{VV}^\mathcal{T}-\mathbf{YY}^\mathcal{T}}_F  \nonumber \\
& = \norm{\mathbf{V}^\mathcal{T}\mathbf{V}}^2_F - 2\norm{\mathbf{V}^\mathcal{T}\mathbf{Y}}^2_F + \norm{\mathbf{Y}^\mathcal{T}\mathbf{Y}}^2_F,
\end{align}

where $F$ denotes the Frobenius norm of the matrix.  In order to remove the contribution of noisy/silent regions in the network training, a voice active detection (VAD) threshold is employed.  Only the embeddings corresponding to time-frequency bins with magnitude greater than the VAD threshold (-40 dB below the maximum amplitude as in \cite{hershey2016deep}) contribute to the above loss calculation.

It should be noted that at the test stage the k-means algorithm is employed to cluster the embeddings using the entire test signal,  thus making the method unsuitable for low-latency processing. In the test stage, the estimated binary masks are applied to the complex spectrogram of the mixture hence mixture phase is utilized. Inverse STFT and overlap-add processing is applied to obtain  separated signals in the time domain. 




\begin{figure}\label{fig: low_latency_block}
\includegraphics[scale=0.5]{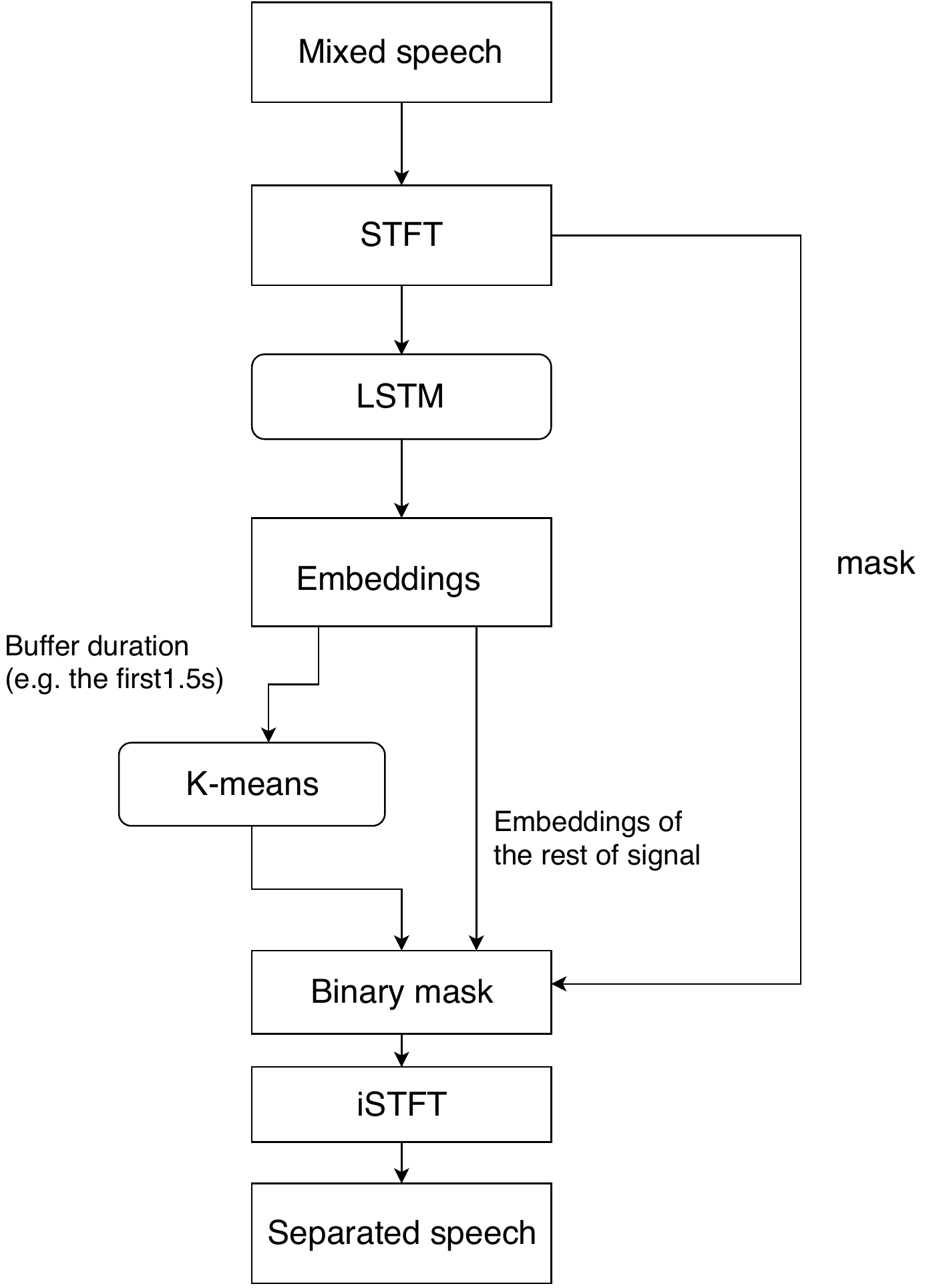}
  \caption{The block diagram of the proposed low-latency deep clustering method.}  \label{fig: low_latency_block}
\end{figure}

\vspace{2ex}

\section{Low-Latency Deep Clustering}\label{sec:low_latency}
In order to make the deep clustering based separation operate with low latency, there are three parts that need to be adapted: \textit{a)} The topology of the neural network is changed from BLSTM as was used in \cite{hershey2016deep} to LSTM in order to produce embedding vectors in an online manner for each frame as they are inputted to the network; \textit{b)} In the baseline method \cite{hershey2016deep}, 32 ms synthesis window length is used. The resulting latency may be prohibitive for certain applications, e.g., hearing aids \cite{agnew2000hearing} as was explained in the introduction. Hence we  shorten the window length to 8 ms; and; \textit{c)} Instead of using the whole signal, we propose using only a certain length in the beginning of the  mixture, which we refer as the buffer, to get the cluster centres. These cluster centres can be then used to  estimate the masks for the rest of the mixture. Please note that since the first few seconds of the signal are used to estimate the cluster centres, the method is not able to separate sources during this initialization stage. However, after the buffer, the rest of the signal will be processed in an online manner. The process of the low-latency version of deep clustering method is depicted in Fig. \ref{fig: low_latency_block} with a buffer duration of 1.5 s. 


\section{EVALUATION} \label{sec:evaluation}
\vspace{1ex}
\subsection{Acoustic material}
The evaluation is done using synthetic two-speaker mixtures generated from Wall Street Journal (WSJ0) corpus. The duration of mixtures is on average around 6 s. The training data set consists of 20,000 two-speaker mixtures created by randomly selecting utterances from 101 different speakers from WSJ0 \textit{si\_tr\_s} that amounts to around 30 hours of training material. Similarly, for the validation data set, we create 5000 two-speaker mixtures that last for around 8 hours having the same speakers as in training data set. The test data is generated from WSJ0 \textit{si\_dt\_05} and \textit{si\_et\_05} and consists of 3000 mixtures and lasts around 5 hours having 18 different speakers. The test data has different speakers from training data and validation data for the purpose of evaluating the separation performance in open conditions as described in \cite{hershey2016deep}.  

We downsample the speech samples from 16 kHz to 8 kHz for reducing the computational requirements and to make the evaluation setup similar to \cite{hershey2016deep}. As the proposed approach (factor \textit{c}) relies on both the speakers being active during the buffer duration, for a fair investigation of the effect of buffer duration and comparison of offline deep clustering to its online counterpart, the same data should be used for evaluation.  Hence in the test set  for  (\textit{c}) we firstly remove the silence from the beginning of both speech signals and sum them to form the mixture thus ensuring  that both speakers are active during the buffer duration ( $\geq$ 100 ms in this work, i.e., all mixtures have both speakers active within at least 100 ms in the beginning). The longer speech signal is trimmed to the length of the shorter utterance before adding to form the mixture. It should also be noted that such test mixtures have a larger degree of overlapped speech and are thus harder to separate.  


\subsection{Metrics}
We use BSS-EVAL toolbox \cite{vincent2006performance} for evaluating the system performance. It consists of three metrics:  signal-to-distortion-ratio (SDR), signal-to-interference-ratio (SIR), and signal-to-artifacts-ratio (SAR). The average SDR of test mixtures without any separation is 0.1 dB.

\begin{table}[b!]
\centering
\caption {Feature and system parameters for offline and online deep clustering experiments.}
\vspace{2ex}
\label{tab:sys_param}
\begin{tabular}{l@{\hspace{3ex}}c@{\hspace{3ex}}c}
\toprule
{} & offline DC & low-latency DC\\ \toprule
Window length & 32 ms & 8 ms\\
Hop length & 8 ms & 4 ms\\
Sequence length & 100 & 200\\
Network   & BLSTM   & LSTM \\
 \midrule
Window  & \multicolumn{2}{c}{Hanning}\\
Sampling frequency & \multicolumn{2}{c}{8 kHz}\\
FFT size & \multicolumn{2}{c}{256}\\
Number of layers & \multicolumn{2}{c}{4}\\
Number of LSTM units & \multicolumn{2}{c}{600}\\
Embedding dimension & \multicolumn{2}{c}{40}\\
\bottomrule
\end{tabular}
\end{table}

\subsection{Experiment setup}
In order to analyze the effect of the following different factors, namely,  \textit{a)} BLSTM vs LSTM networks,  \textit{b)} 32 ms vs 8 ms window length, and, \textit{c)} different buffer duration for low-latency process, we conduct separate experiment for each of these.

The baseline framework is taken to be the one used in \cite{hershey2016deep}. It consists of a BLSTM network with four layers and 600 units in each layer followed by a time-distributed dense layer. The number of units in the time distributed dense layer is the product of the number of embedding dimensions and the number of effective FFT points. Hyperbolic tangent (tanh) is used as the activation function in this layer. After the dense layer, L2 normalization is used to bound the embedding vectors to unit norm. The same parameters have been used for the LSTM network in order to analyze the effect of factor \textit{a)}. To compare the effect of different window length, the same LSTM network and a shorter window length  of 8 ms  for STFT feature extraction are used. For a fair comparison with the baseline, the network must be trained with the sequences having the same time context. The baseline BLSTM was trained on 100 frame sequences (800 ms). Here we reduce the hop length to 4 ms hence the sequence length is increased to 200 (800 ms). We first train the network for 100 frame sequences and then after convergence continue training with 200 frame sequences, known as curriculum learning used in \cite{Isik+2016} and first introduced in \cite{bengio2009curriculum}. The idea is to pre-train a network on an easier task first improves learning and generalization. Finally,  \textit{c)} is studied by varying the buffer duration using the network with the same LSTM network (four layers and 600 units in each layer) with 8 ms window length. The same FFT size (256) is used for both offline (32 and 8 ms frame length) and online deep clustering (8 ms frame length) frameworks, i.e., zero padding is used wherever required.


During the training process, the 'Adam' optimizer is utilized \cite{kingma2014adam}. The  Keras \cite{chollet2016keras}  and Tensorflow \cite{45381} libraries are used for network training, and Librosa library \cite{librosa} is used for feature extraction and signal reconstruction in this paper. In order to reduce overfitting, early stopping  method is used \cite{giles2001overfitting} by monitoring the loss on validation data and stopping the training when no decrease in it is observed for 30 epochs. The embedding dimension is set to 40 and the VAD threshold is 40 dB, similar to the original study \cite{hershey2016deep}. The detailed description of the parameters for the network can be found from Table \ref{tab:sys_param}.

It should be noted that Fig. \ref{fig: low_latency_block} depicts the real world use case where a buffer duration in the beginning of an utterance is used for estimating clusters and the  separation starts after that.  This however makes acoustic material used for evaluation dependent upon the buffer duration if the same utterance is used for both cluster estimation and evaluation. To deal with this mismatch, for each test utterance, we randomly select another utterance (cluster utterance) belonging to the same speaker pair and use it to estimate clusters. Different buffer lengths can thus be sampled from the beginning of this cluster utterance for the same test utterance in order to study the effect of factor \textit{c}. Moreover, the VAD threshold is used during cluster estimation to exclude the effect of noisy time-frequency bins.



\begin{figure}[t!]
\centering
\includegraphics[scale = 0.6,clip]{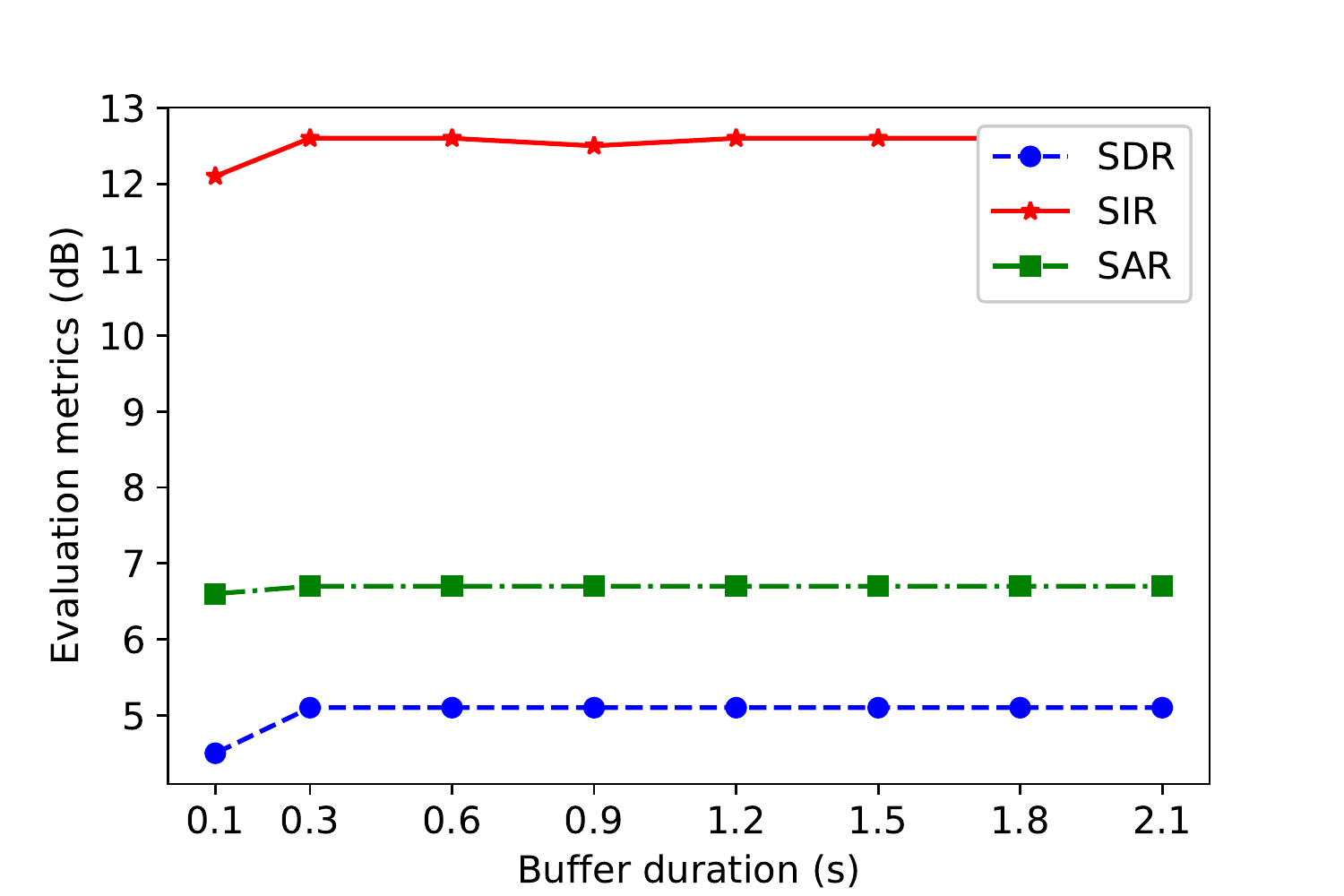}\\
  \caption{Evaluation metrics (dB) for LSTM network with  different buffer durations (factor \textit{c} in the experimental set up.)}
  \label{fig:low_latency_graph}
\end{figure}

\subsection{Results and discussion}
We calculate the mean of evaluation metrics, SDR, SIR, and SAR over all the test mixtures. All the test mixtures are formed such that both constituent speech signals are active within the lowest buffer duration used in experiments (100 ms here). Hence they have a higher overlap between the constituent speech. As described  in the previous section, we adopt the strategy of estimating clusters on different utterances than test utterances. Hence  the same dataset can be used for evaluation for both offline and online deep clustering methods. The results with baseline offline case is shown in Table \ref{tab:offline_result}. Similarly, the evaluation metrics corresponding to the LSTM network with 32 and 8 ms window lengths is shown as well.  The online LSTM in Table \ref{tab:offline_result}  refers to low-latency LSTM with 8 ms window length and 1.5 s buffer time.  It can be observed  that the separation performance is one dB lower in terms of SDR  while replacing BLSTM to LSTM while keeping the same window length. Moreover, by decreasing the window length to 8 ms, SDR  degrades by about 2.1 dB as compared to the baseline.


The effect of varying buffer duration on performance metrics is shown in \mbox{Fig. \ref{fig:low_latency_graph}}. Two interesting observations can be made from it: firstly, even with a short buffer duration, e.g., 100 ms, relatively reasonable separation performance can still be achieved (4.5 dB); and secondly after a certain buffer length more information does not lead to a drastic improvement in separation performance. This means a small buffer duration, even as low as 300 ms, can yield good separation provided both the constituent speakers are active during it.




\begin{table}[!t] 
\centering
\caption {Evaluation metrics (dB) of different variants of the offline method and the online method with 1.5s buffer. Here online refers to factor \textit{c} in the experimental setup}
\label{tab:offline_result}
\vspace{2ex}
\vfill
\begin{tabular}{lcccc}
\toprule
 {} & \hspace{0mm}Window length &\hspace{0mm} SDR & \hspace{0mm}SIR &\hspace{0mm} SAR\\
 \midrule
 BLSTM & 32  ms & 7.9  & 15.6 & 9.2\\
LSTM & 32 ms & 6.9  & 14.5 & 8.4\\
LSTM & 8 ms & 5.8  & 13.6 & 7.2\\
\midrule
Online LSTM & 8 ms & 5.1  & 12.6 & 6.7\\
 (1.5s buffer)& &  & \\
\bottomrule
\end{tabular}
\end{table}

\section{CONCLUSION} \label{sec:conclusion}
The paper proposes a low-latency adaptation of deep clustering based speech separation. In particular, a buffer signal duration in the beginning of audio mixture is used for estimating cluster centres corresponding to the speakers present in the mixture. This duration serves as an 'initialization' period after which the rest of the speech mixture  is  processed in online manner. Moreover, separation performance of the method using an LSTM network and a short synthesis window length of 8 ms, as required by real-time operation, has been studied. A degradation in SDR of about one dB is observed for the former and 2.1 dB for the latter as compared to the baseline. Finally, we investigate how the buffer duration affects the separation result and observe that even very short buffer duration, e.g. 300 ms, is sufficient to estimate clusters for reasonable separation.


\bibliographystyle{IEEEtran}
\small

\end{document}